\documentclass[journal=jacsat,manuscript=article]{achemso}
\usepackage[version=3]{mhchem} 
\usepackage{siunitx}
\usepackage[dvipsnames]{xcolor}
\usepackage{amssymb}
\usepackage{pifont}
\usepackage{multirow}
\usepackage{subfig}
\usepackage{wrapfig}
\usepackage{hyperref}
\usepackage{graphicx}

\usepackage{amsmath} % or simply amstext

\usepackage{tabularx,colortbl}
\usepackage{capt-of}
\usepackage{adjustbox}
\usepackage{tabularx}
\usepackage[font=normalsize,labelfont={bf},font={bf}]{caption}
\usepackage{caption}
\author{Akshat Chaudhari}
\affiliation[mse]
{Department of Material Science and Engineering, Carnegie Mellon University, 15213, USA}

\author{Chakradhar Guntuboina}
\affiliation[ece]
{Department of Electrical and Computer Engineering, Carnegie Mellon University, 15213, USA}

\author{Hongshuo Huang}
\affiliation[mse]
{Department of Material Science and Engineering, Carnegie Mellon University, 15213, USA}

\author{Amir Barati Farimani}
\email{barati@cmu.edu}
\affiliation[meche]
{Department of Mechanical Engineering, Carnegie Mellon University, 15213, USA}
\alsoaffiliation[biomed]
{Department of Biomedical Engineering, Carnegie Mellon University, 15213, USA}
\alsoaffiliation[mld]
{Machine Learning Department, Carnegie Mellon University, 15213, USA}

\title[An \textsf{achemso} demo]
{AlloyBERT: Alloy Property Prediction with Large Language Models}

\begin{document}

\begin{abstract}
\noindent
The pursuit of novel alloys tailored to specific requirements poses significant challenges for researchers in the field. This underscores the importance of developing predictive techniques for essential physical properties of alloys based on their chemical composition and processing parameters. This study introduces AlloyBERT, a transformer encoder-based model designed to predict properties such as elastic modulus and yield strength of alloys using textual inputs. Leveraging the pre-trained RoBERTa encoder model as its foundation, AlloyBERT employs self-attention mechanisms to establish meaningful relationships between words, enabling it to interpret human-readable input and predict target alloy properties. By combining a tokenizer trained on our textual data and a RoBERTa encoder pre-trained and fine-tuned for this specific task, we achieved a mean squared error (MSE) of 0.00015 on the Multi Principal Elemental Alloys (MPEA) data set and 0.00611 on the Refractory Alloy Yield Strength (RAYS) dataset. This surpasses the performance of shallow models, which achieved a best-case MSE of 0.00025 and 0.0076 on the MPEA and RAYS datasets respectively. Our results highlight the potential of language models in material science and establish a foundational framework for text-based prediction of alloy properties that does not rely on complex underlying representations, calculations, or simulations. 
\end{abstract}

% \clearpage
\section{1. Introduction}
Discovering novel alloys for improved performance, efficiency, and innovation is a constant demand across numerous industries like energy, medical and healthcare, aerospace, and robotics to name a few, in order to address pressing difficulties, optimize functionality, and unveil possibilities for innovative applications. The search for novel alloys with unique properties tailored for specific needs presents a wide range of challenges. This necessitates the need of methods to determine important physical properties of alloys based on their chemical and processing information. Since alloys are mixtures of metals and non-metals, the sheer number of potential combinations, often with different processing and manufacturing techniques makes experimental exploration very slow and impractical. This vast space containing multitude of possible combinations necessitates robust computational tools capable of predicting physical properties from atomic configurations. These include Density Functional Theory (DFT) calculations which can help provide understanding relationship between composition and properties to tailor alloys with specific properties in mind\cite{dftmaterialproperties}\cite{dftalloy}\cite{dftmaterialsresearch}. However, alloys in general are large and complex systems which can render the DFT calculations to be computationally extensive which limits their usage for the vast alloy spaces\cite{dftcomputationallyexpensive}. Certain material properties, like multiscale phenomena, emergent behaviour, and temperature dependence, cannot be fully captured by the local density approximation inherent in DFT\cite{dftlimits}. This requires incorporating additional methods or higher-level theoretical frameworks to account for these complexities. DFT also relies on various approximations, such as exchange-correlation functionals, which can introduce inaccuracies in certain cases\cite{dftfunctionalinaccuracies}\cite{dftexchangecorrelation}.\\
Traditional machine learning methods like linear regression, support vector machines, random forests, neural networks, etc have been found to be cost effective and successful at accelerated discovery of candidate alloys from the wide range of possible combinations\cite{vasudevan2021machine} \cite{hart2021machine}. They are shown to be able to handle non-linear relationship between multiple variables which sometimes proves to be difficult for traditional methods\cite{mlnonlinear}. They can serve as a cost-effective tool for initial property assessment, reducing the need for time-consuming and expensive laboratory experiments and simulations\cite{moformer}. Moreover, recent advancements in graph representations and Graph Neural Networks (GNNs) have shown promising results in the field of chemistry, enabling the modeling of molecular structures and interactions with high accuracy and efficiency\cite{gnnmolecules}\cite{crystalgnn}. These techniques offer a novel approach to understanding complex molecular relationships, complementing traditional machine learning methods and enhancing the discovery process of novel alloys as they show the role of representation in property prediction tasks with machine learning models. However, certain models like neural networks can act like “black boxes” which makes it difficult to understand the reasoning behind the model predictions and sometimes the results may not generalize well with new data and alloys outside the scope of the training data\cite{nnblackbox}. Training and validating ML models require high-quality, comprehensive datasets, which can be limited in the field of alloy research. A lot of times the experimental data is very messy and this is something that traditional machine learning algorithms like linear regression and random forests don’t deal with very well. There has to be a specific format in which data needs to be fed to the model and all input fields must have valid values. Data may not always be available in this format which puts restriction on how ML algorithms can be used. The development of generative language models has led to increased interest in generating text based descriptions for molecules in chemistry\cite{textmolecules} and this presents a prospect to representing alloys and their properties similarly which can be better interpreted by humans\cite{textmaterials}. When an alloy is represented in textual format, it can be passed into a deep neural network, however breaking the input into meaningful tokens will present a major challenge as strategically placing important tokens throughout the text's potentially long span can complicate the design of neural network models.  \\
A variety of Transformer-based models, including BERT\cite{bert} \cite{bertelmo}, RoBERTa\cite{roberta}, GPT\cite{gptroberta}, ELMo\cite{bertelmo}, and LLAMA\cite{llama}, have demonstrated remarkable performance on a variety of natural language processing (NLP) tasks in recent years. The Transformer's unique strength is its deft use of an attention mechanism, which enables the model to recognize meaningful relationships between sentence tokens without depending just on hidden states from the past\cite{attentionisallyouneed}. There have been attempts in the field of materials engineering and chemistry to put the power of transformer models to use for various applications\cite{multimodalgraph, catberta, transpolymer, peptidebert, gpcrbert, matinformer, moformer}. For example, TransPolymer\cite{transpolymer} employs polymer sequence representations for predicting polymer properties, leveraging pretraining through masked language modeling (MLM) on unlabeled data. Similarly, CatBERTa\cite{catberta} predicts adsorption energy of adsorbate from textual descriptions of the catalyst\cite{catberta}. Transformers have also been shown to be effective in the biomedical domain with applications such as PeptideBERT\cite{peptidebert} and GPCR-BERT\cite{gpcrbert}. PeptideBERT is a protein language model specifically designed for predicting essential peptide properties like hemolysis and nonfouling using the pretrained ProtBert model. Through fine-tuning, it achieves state-of-the-art results, showcasing remarkable performance on shorter sequences. GPCR-BERT, on the other hand, uses language model techniques to understand the sequential design of G protein-coupled receptors (GPCRs) by analyzing attention weights and hidden states. This leads to insights into residue contributions and relationships with conserved motifs, ultimately resulting in high accuracy in predicting hidden residues within the motifs and explaining higher-order interactions within their conformations. \\
We propose the AlloyBERT model which leverages transformer models’ unique abilities to interpret textual data, which is designed to predict different alloy properties such as elastic modulus and yield strength. We aim to show the abilities of transformer-based machine learning models to make sense of human readable data to find relevant relations between provided information to output the above-mentioned alloy properties. This approach opens a new avenue for property prediction in alloys and can serve as a valuable tool to get critical properties for alloys bypassing the computational expensive DFT calculations and since this method takes in simple text input, it can serve as a very practical tool as data can be very messy in experimental settings and simply inputting in known and observed properties of a particular alloy can be used to give accurate predictions about its physical properties.

\begin{figure}[h]
    \centering
    \includegraphics[width=1\textwidth]{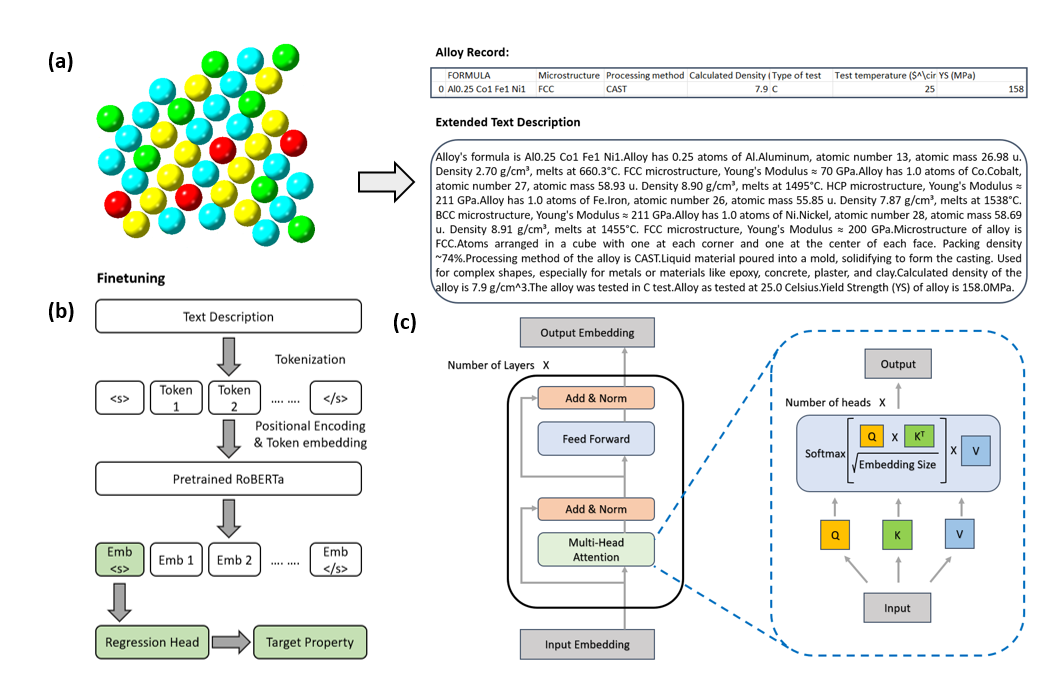}
    \caption{Overview of AlloyBERT: (a) Known properties of alloy are converted to elaborate textual description containing additional information about the constituents, processing and other physical properties. (b) Visualization of finetuning process. The embedding from the special token ‘\textless s\textgreater’ is input to the regression head, comprising a linear layer and activation layer. (c) Illustration of the Transformer encoder and multi-head attention mechanism.} 
    \label{fig:example}
\end{figure}

\section{2. Methods}
\subsection{2.1. Model Architecture}

Our model architecture is built upon the RoBERTa\cite{roberta} (Robustly optimized BERT pretraining approach) model, a variant of BERT\cite{bert} (Bidirectional Encoder Representations from Transformers) that uses a different pretraining approach and has been shown to outperform BERT on several benchmarks. 
RoBERTa, like BERT, is a transformer-based model. The transformer architecture is based on self-attention mechanisms and consists of an encoder and a decoder. However, in RoBERTa and BERT, only the encoder part is used. The encoder is composed of a stack of identical layers, each with two sub-layers: a multi-head self-attention mechanism and a position-wise fully connected feed-forward network.\\
The self-attention mechanism\cite{attentionisallyouneed}, also known as scaled dot-product attention, is a key component of the transformer architecture. It allows the model to weigh the importance of words in the input sequence when encoding a specific word. This is particularly useful for understanding the context and handling long-range dependencies in the text.\\
In the self-attention mechanism, each word in the input sequence is associated with a query (Q), a key (K), and a value (V). The queries and keys are used to compute attention scores, which determine how much each word should contribute to the encoding of a given word. The values are then weighted by these attention scores and summed to produce the final output. The attention score between a query and a key is computed as the dot product of the query and key, divided by the square root of the dimension of the key. This can be represented as: $$Attention(Q,K,V) = softmax(\frac{QK^T}{\sqrt(d_k)})V$$

where $d_k$ is the dimension of the key.

The multi-head attention mechanism\cite{attentionisallyouneed} is a variant of the self-attention mechanism that applies the self-attention operation multiple times in parallel. The outputs of these parallel operations are then concatenated and linearly transformed to produce the final output. This allows the model to capture different types of information from the input sequence. The position-wise feed-forward network consists of two linear transformations with a ReLU activation in between. RoBERTa differs from BERT in terms of its pretraining approach. While BERT uses a masked language model and next sentence prediction tasks for pretraining, RoBERTa only uses the masked language model task with dynamic masking, and it trains with much larger batch sizes and byte pair encoding\cite{bpe}, which results in improved performance\cite{roberta}.\\
In the pretraining stage, we utilize the RoBERTa model for masked language modeling. We use a pretrained tokenizer to prepare our text corpus for this stage. The model is trained to understand the semantics of the text and generate meaningful sentence embeddings.

\subsection{2.2. Datasets}
 For this study we utilized two primary datasets to investigate how transformer models perform for predicting target property values with text input against shallow machine learning models.
 \begin{enumerate}
     \item Multi Principal Elemental Alloys (MPEA) dataset: This dataset has been obtained from Citrine Informatics and it contains mechanical properties of several alloys and the target value of experimental Young’s modulus for this dataset has been used. The full dataset has 1546 entries.

     \item Refractory Alloy Yield Strength (RAYS) dataset: The dataset contains experimental yield strength for refractory alloy. The full dataset has 813 entries with alloy composition, testing temperature from previous literature and MPEA\cite{borg2020expanded,gorsse2018database,couzinie2018comprehensive} dataset. The dataset presents average yield strength values derived from different processing method.
 \end{enumerate}

\subsection{2.3. Data Preprocessing}
The Citrine Informatics MPEA dataset required preprocessing for effective evaluation with shallow machine learning models, allowing for a robust comparison with AlloyBERT. Irrelevant columns related to data provenance, such as doi, year, title, and reference, were excluded. Due to the dataset's limited size of 729 usable entries, columns with fewer than 300 valid values and those containing highly similar entries were removed to prevent overfitting and improve model generalizability. The MPEA dataset was further processed when used with shallow machine learning models, namely, features of type `string' were transformed into 1-hot encodings. Additionally, the chemical formulas of the alloys were parsed, and new columns were added to represent the elemental composition of each alloy. No cleaning was necessary for the refractory alloy yield strength dataset. These preprocessing steps prepared the MPEA dataset for effective training and evaluation with shallow models, enabling a fair and informative comparison of their performance with AlloyBERT. The datasets were then converted into textual descriptions, to do this, we utilized a Python script to parse dataset records into their features, augmenting them with additional relevant information about elemental composition and other properties.

\subsection{2.4. Training Procedure}
Textual descriptions were generated for the alloys in the dataset using a Python script. The chemical formulas of the alloys were employed to create compositional statements, supplemented by general elemental properties such as atomic mass and number, as well as density. Additional details pertaining to processing methods like casting and wrought, as well as microstructural characteristics like BCC (body-centered cubic) and FCC (face-centered cubic), were incorporated into the textual descriptions. The text descriptions we generated were comprehensive encapsulating details from atomic level to microstructural level, and this is very important because the performance of downstream tasks in pretraining models is heavily influenced by both the quality and quantity of the pretraining data\cite{pretraindataquality}. Table\ref{tab:my_label} exemplifies a few records from the MPEA dataset.

\begin{table}[H]
\small
    \begin{tabularx}{\textwidth}{ |>{\centering\arraybackslash}p{3cm}|>{\centering\arraybackslash}p{2.8cm}|>{\centering\arraybackslash}p{2cm}|>{\centering\arraybackslash}p{1.3cm}|>{\centering\arraybackslash}p{0.8cm}|>{\centering\arraybackslash}p{1.74cm}|>{\centering\arraybackslash}p{1.8cm}| } 
    \hline
    \textbf{Formula} & \textbf{Microstructure} & \textbf{Processing} & \textbf{$\rho$ $(\frac{g}{cm^3})$} & \textbf{Test} & \textbf{Temp($^\circ$C)} & \textbf{YS(MPa)}\\ \hline
    Al0.25Co1Fe1Ni1 & FCC & CAST & 7.9 & C & 25 & 158 \\
    Al1Cr1Mo1Nb1Ti1 & BCC & CAST & 6.6 & C & 600 & 1060
\\
    Hf1Nb1Ta1Ti1 & BCC & ANNEAL & 10.9 & C & 20 & 860 \\
    Fe1Mn1Ni1 & FCC & WROUGHT & 8.1 & C & 25 & 460
    \\
    \hline
    \end{tabularx}
    \caption{\label{tab:my_label}Some data records from MPEA dataset}
\end{table}

Generated textual Description for the first record in Table \ref{tab:combined}: Alloy's formula is Al0.25 Co1 Fe1 Ni1.Alloy has 0.25 atoms of Al.Aluminum, atomic number 13, atomic mass 26.98 u. Density 2.70 $g/cm^3$, melts at 660.3°C. FCC microstructure, Young's Modulus 70 GPa.Alloy has 1.0 atoms of Co.Cobalt, atomic number 27, atomic mass 58.93 u. Density 8.90 g/cm³, melts at 1495°C. HCP microstructure, Young's Modulus 211 GPa.Alloy has 1.0 atoms of Fe.Iron, atomic number 26, atomic mass 55.85 u. Density 7.87 g/cm³, melts at 1538°C. BCC microstructure, Young's Modulus 211 GPa.Alloy has 1.0 atoms of Ni.Nickel, atomic number 28, atomic mass 58.69 u. Density 8.91 g/cm³, melts at 1455°C. FCC microstructure, Young's Modulus 200 GPa.Microstructure of alloy is FCC.Atoms arranged in a cube with one at each corner and one at the center of each face. Packing density ~74\%.Processing method of the alloy is CAST.Liquid material poured into a mold, solidifying to form the casting. Used for complex shapes, especially for metals or materials like epoxy, concrete, plaster, and clay.Calculated density of the alloy is 7.9 g/cm³. The alloy was tested in C test.Alloy as tested at 25.0 Celsius.Yield Strength (YS) of alloy is 158.0MPa.\\

A Byte Pair Encoding (BPE)\cite{bpe} tokenizer was trained on the textual data seperately for each dataset. This tokenizer iteratively merges frequent contiguous byte pairs (character pairs) into new tokens, thereby constructing a vocabulary that adapts to the language patterns present in the data. BPE possesses the capability to tokenize any word, even if it was not encountered during training, by breaking it down into subwords. This feature is particularly valuable for handling diverse text in real-world scenarios. Moreover, BPE's subword tokens often result in smaller vocabularies compared to those generated by word-based tokenizers\cite{bpesmallervocab}. Consequently, this reduction in vocabulary size contributes to a smaller model size, decreased memory requirements, and reduced training time, rendering it suitable for large-scale pretraining endeavors.\\
Subsequently, the RoBERTa model was employed in conjunction with the tokenizer to conduct pre-training through masked language modeling (MLM) on the textual data. This model encompasses a stack of Transformer encoder layers. These layers are responsible for processing the input text sequence and deriving contextual representations. Moreover, a dedicated MLM head is added atop the pre-trained model, which leverages the encoder outputs to predict the masked words within the input sequence. During the training process, a fraction of the input tokens (e.g., 15\%) are randomly masked and substituted with a special masking token or random tokens. Subsequently, the MLM head utilizes the hidden representations from the final encoder layer to estimate the probabilities of the original words for each masked token. Notably, this approach improves upon BERT by utilizing dynamic masking instead of the static masking employed in BERT, resulting in slightly different learning dynamics.\\
Following the completion of the MLM phase, the pre-trained model is loaded, and a regression head is affixed to it to facilitate the prediction of target values for the alloy. The model undergoes training for 50 epochs, utilizing a batch size of 1 and a learning rate of 1.0e-5 with the AdamW\cite{adamw}\cite{adam} optimizer. Additionally, a linear learning rate scheduler is employed to gradually decrease the learning rate from the initial value to 0 over a specified number of training steps. We have used Mean Squared Error (MSE) as our loss function.  The training process was executed on an NVIDIA GeForce RTX 3050 Ti GPU.

\section{3. Results and Discussion}

We compared the performance of our model against a suite of shallow models using the Mean Squared Error (MSE) metric. The performance of the shallow models is depicted in Table \ref{tab:shallow-performance}. We evaluated the performances of linear regression\cite{linearregression}, random forests\cite{randomforests}, support vector regression\cite{svm}\cite{svr} and gradient boosting\cite{gradientboosting} algorithms on the two datasets. Gradient boosting delivered the best results amongst these 4 methods on the MPEA dataset with a MSE of 0.00025 and random forests acheived the lowest MSE on the RAYS dataset of 0.0076.

The 4 different types of textual description formats we experimented with are outlined in \ref{tab:performance}. Initially we passed the information for each record as a string of numbers from the dataset as it is. Then we converted this into english sentences with additional information about some features like the processing methods (casting, wrought, etc) and microstructure (BCC, FCC, etc). Then we extracted elemental composition from each alloy and added atomic number and masses of these elements to the text input of that alloy. Finally, we added information about physical properties of these elements as well like the melting and boiling points, Young's modulus of the individual elements, their microstructural information.

The performance of our model on these 4 types of inputs is presented in \ref{tab:combined}. `Finetune only' involved simply using the Roberta base model for prediction. `Pretrain + Finetune' involved using a custom Byte Pair Encoding (BPE) tokenizer (trained on our text corpus from the dataset) and masked language modelling (MLM) on the Roberta model prior to finetuning.

\begin{figure}[H]
    \centering
    \includegraphics[width=1\textwidth]{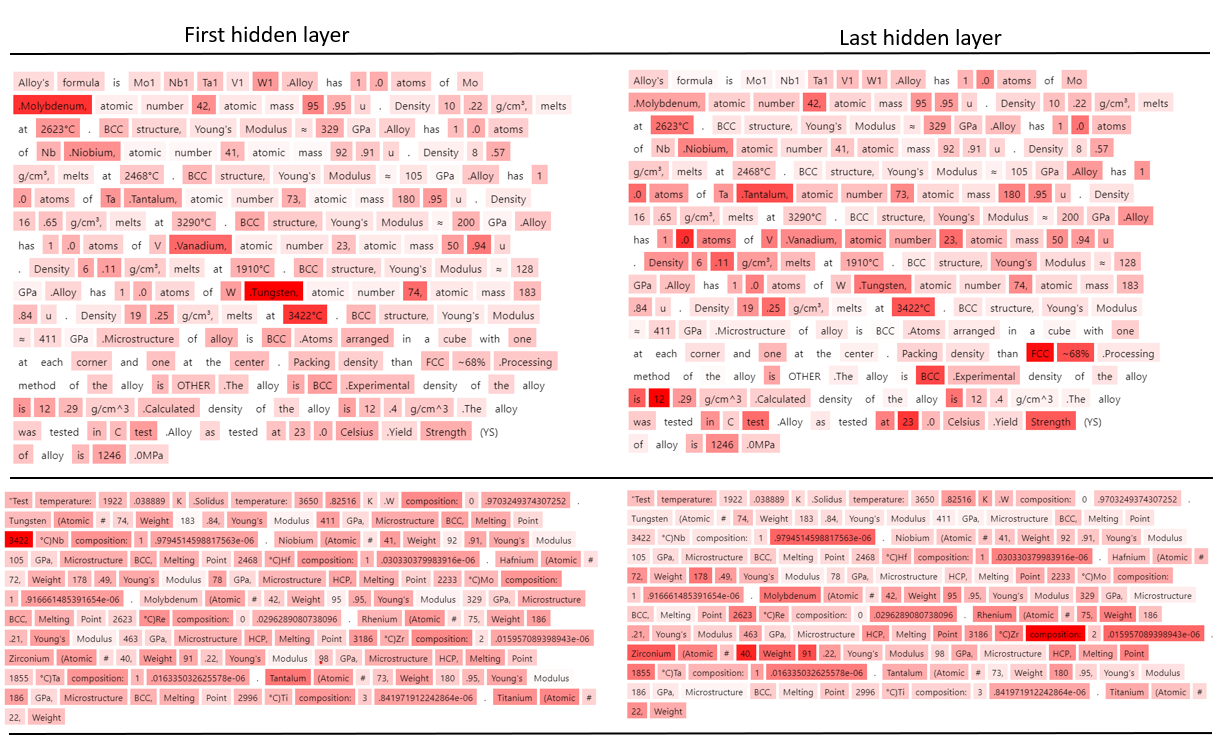}
    \caption{Attention scores visualization\cite{selfattention} from AlloyBERT. Left column shows attention scores from the initial hidden layer and the right column shows attention scores from the final hidden layer. Top row is a String 4 text format from MPEA dataset and bottom row is a String 4 representation of input from RAYS dataset.} 
    \label{fig:attn}
\end{figure}

The self-attention scores diagram\cite{selfattention} (Fig. \ref{fig:attn}) provides a visual representation of the attention mechanism used by the RoBERTa transformer model for the initial and final hidden layers. In the context of our study, this visualization highlights words in the input text that the model focuses on when making predictions about alloy properties. The intensity of the highlighting corresponds to the attention score assigned to each word, indicating the importance of that word in the prediction process.

\begin{table}[H]
    \centering
    \begin{tabular}{|c|c|c|c|c|}
        \hline
        Model & MPEA Dataset & RAYS Dataset\\
        \hline
        Linear Regression & 0.0004 & 0.0111  \\
        Random Forest & 0.0005 & 0.0076  \\
        SVR & 0.0248 & 0.0117  \\
        Gradient Boosting & 0.00025 & 0.0087  \\
        \hline
    \end{tabular}
    \caption{Performance (MSE) for Shallow Models}
    \label{tab:shallow-performance}
\end{table}

\begin{table}[H]
    \centering
    \begin{tabular}{|l|p{0.8\linewidth}|}
        \hline
        No. & Information \\
        \hline
         String 1 & Basic Featurization: Input features (numbers) from original dataset passed in string format \\
        String 2 & Dataset features passed as sentences with added information of features (processing, microstructure) \\
        String 3 & String 2 + Basic information of elements in alloy (atomic number, atomic mass) \\
        String 4 & String 3 + Additional information about physical properties of the elements (melting point, boiling point, Youngs modulus)\\     
        \hline
    \end{tabular}
    \caption{Types of input formats}
    \label{tab:performance}
\end{table}

\begin{table}[htbp]
    \centering
    \begin{tabular}{|c|c|c|c|c|}
        \hline
        \textbf{String Type} & \textbf{Training Method} & \textbf{MPEA Dataset} & \textbf{RAYS Dataset}\\
        \hline
        \multirow{2}{*}{String 1} & Finetune only & 0.0018 & 0.0101  \\
         & Pretrain + Finetune & 0.00349  & 0.00789  \\
        \hline
        \multirow{2}{*}{String 2} & Finetune only & 0.00536 & 0.00839  \\
         & Pretrain + Finetune & 0.00668  & 0.00766  \\
        \hline
        \multirow{2}{*}{String 3} & Finetune only & 0.00325 & 0.00799  \\
         & Pretrain + Finetune & 0.00481  & 0.00626  \\
        \hline
        \multirow{2}{*}{String 4} & Finetune only & 0.00029 & \textbf{0.00611}  \\
         & Pretrain + Finetune & \textbf{0.00015}  & 0.007  \\
        \hline
    \end{tabular}
    \caption{Performance of AlloyBERT using different training methods}
    \label{tab:combined}
\end{table}

\begin{figure}%
    \centering
    \subfloat[\centering MPEA]{{\includegraphics[width=7cm]{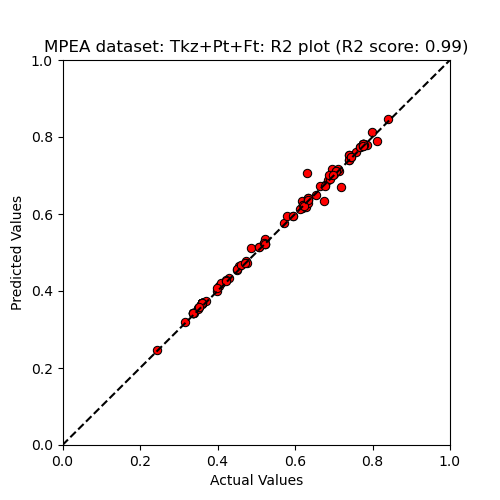} }}%
    \qquad
    \subfloat[\centering RAYS]{{\includegraphics[width=7cm]{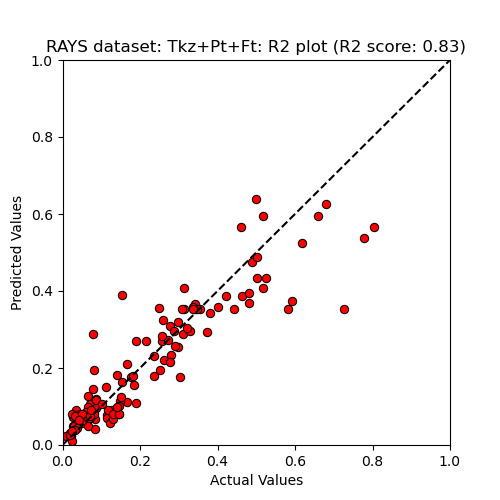} }}%
    \caption{Parity plots for AlloyBERT predictions. X axis corresponds to actual values of elastic modulus in MPEA dataset and yield strength in RAYS dataset and Y axis corresponds to values predicted by AlloyBERT}%
    \label{fig:parity}%
\end{figure}

For the initial 2 types of text inputs the MSE for MPEA dataset increased as we added more text information which was opposite to how we expected a transformer model to perform. For String 3, MSE decreases significantly when using a tokenizer. However, for String 4, where the input data is most elaborate, using custom trained tokenizer along with pretraining and finetuning achieves minimum MSE of 0.00015 off all methods and text formats for the MPEA dataset.

In the RAYS dataset, for String 1 there is a significant drop in the MSE from finetuning only to pretraining and finetuning method. For rest of input formats, the MSE does not change considerably, however the best MSE for this dataset is again achieved when using the most elaborate string description using Roberta base model.

Parity plots for both the datasets are shown in Fig. \ref{fig:parity}. We achieved an $R^2$ score of 0.99 on the MPEA dataset indicating a very strong correlation between the predicted and actual alloy properties, suggesting that the model captures the underlying patterns in the data exceptionally well. For the RAYS dataset, we achieved an $R^2$ score of 0.83 indicating a strong correlation.

\section{4. Conclusion}
This work demonstrates the effectiveness of transformer models in the field of alloy property prediction, particularly when using human-interpretable textual inputs. Despite initial observations where increasing text information led to unexpected MSE behavior in the MPEA dataset, further analysis revealed that the most elaborate textual descriptions, when combined with a custom-trained tokenizer and both pretraining and finetuning, resulted in the minimum MSE of 0.00015 across all methods and text formats for this dataset. Similarly, in the RAYS dataset, the most elaborate string description using the RoBERTa base model consistently yielded the best MSE of 0.00611.

In addition, we demonstrated that the Pretrain + Finetune method achieves the lowest MSE in the MPEA dataset and a significant drop in MSE observed in the RAYS dataset when moving from Finetune only to Pretrain + Finetune. This highlights the importance of utilizing transformer models alongside comprehensive textual inputs and custom tokenizers for accurate alloy property predictions.

Furthermore, the high $R^2$ scores of 0.99 on the MPEA dataset and 0.83 on the RAYS dataset indicate the strong predictive power of our AlloyBERT model. Overall, these findings suggest that transformer models, when coupled with human-interpretable textual inputs, can be a valuable tool in advancing the field of alloy property prediction.

\section{5. Data and Software Availability}
The necessary code used in this study can be accessed here: \url{https://github.com/cakshat/AlloyBERT}

\begin{acknowledgement}
This work is supported by a start-up fund from the Mechanical Engineering Department at Carnegie Mellon University.
\end{acknowledgement}

\bibliography{reference}

\end{document}